\documentclass{ws-procs9x6}

\begin{document}

\newcommand{\lsim}{\buildrel < \over {_\sim}}
\newcommand{\gsim}{\buildrel > \over {_\sim}}
\def\sinhat{\hat{s}^2}

\title{Chiral Symmetries and Low Energy Searches\\ for New Physics}

\author{Michael J. Ramsey-Musolf$^*$ }

\address{Kellogg Radiation Laboratory, California Institute of Technology\\
Pasadena, CA 01125 U.S.A\\
{\rm and}\\
Department of Physics, University of Wisconsin-Madison\\
Madison, WI 53706 U.S.A.\\
$^*$E-mail: mjrm@caltech.edu\\
www.its.caltech.edu/$\sim$mjrm/}

\begin{abstract}
I discuss low energy searches for new physics beyond the Standard Model, identifying the role played by chiral symmetries in these searches and in various new physics scenarios. I focus in particular on electric dipole moment searches; precision studies of weak decays and electron scattering; and neutrino properties and interactions.
\end{abstract}

\keywords{Standard Model, Electroweak Interactions, Chiral Symmetry}

\bodymatter

\section{Introduction}
\label{sec:intro}
The search for physics beyond the Standard Model (BSM) lies at the forefront of the intersection of nuclear physics with particle physics and cosmology. In this talk, I attempt to give an overview of low-energy studies that are being used in this search and try to describe ways in which they complement present and future high energy collider studies. As theme for this meeting is the broken chiral symmetry of QCD, I will endeavor to highlight the role played by chiral symmetries in both the low energy BSM searches and various BSM scenarios. In particular, I will address four questions:
\begin{itemize}

\item[i)] What were the fundamental symmetries that governed the microphysics of the early universe?

\item[ii)] Were there additional (broken) chiral symmetries?

\item[iii)] What insights can precision low energy ($E << M_Z$) studies provide? 

\item[iv)] How does the approximate chiral symmetry of QCD affect the low energy search for new symmetries?

\end{itemize}

As I have described elsewhere\cite{Ramsey-Musolf:2006ur,Ramsey-Musolf:2006ik}, in thinking about fundamental symmetries and the microphysics of the early universe, I like to break cosmic history into three periods: (A) the era of broken Standard Model symmetry, starting from the moment when electroweak symmetry-breaking (EWSB) occurred until the present; (B) the era from the Big Bang until EWSB; and (C) the brief period of EWSB itself. The broken symmetries of the SM -- including the approximate SU(3)$_L\times$SU(3)$_R$ chiral symmetry of strong dynamics involving light quarks -- provide a remarkably successful framework for explaining many phenomena of the present universe, such as the abundance of light elements, weak interactions in stars, and the chiral dynamics of pions and nucleons. Of course, there remain many important but poorly understood aspects of SM dynamics, such as the mechanism for confinement and the possible deconfinement-confinement and chiral symmetry breaking ($\chi$SB) phase transitions that the universe experienced subsequent to EWSB. By and large, however, the SM represents a triumph of 20th century physics with its simple, symmetry-based framework for explaining so much of what we observe today.

Where the SM starts to run into shaky ground begins with EWSB. We believe that the electroweak symmetry of the SM broke down to that of electromagnetism through the Higgs mechanism, whereby elementary particles received non-vanishing mass proportional to the vacuum expectation value (vev) of the neutral Higgs field: $m\propto\langle H^0\rangle\equiv v/\sqrt{2}$. One should emphasize, however, that we have no direct experimental proof that this idea is correct, as no Higgs particle has yet been observed. Direct searches at LEP II place a lower bound of about 114 GeV on its mass, whereas global analyses of precision electroweak data suggest that the mass of the SM Higgs should be less than 166 GeV \cite{lepewwg07}. However, it may be that the SM picture of EWSB is too simple. There may be more than one Higgs field, as in supersymmetric models, for example, or EWSB may not even occur through the Higgs mechanism at all. One hopes that the searches for the Higgs boson at the Tevatron and LHC will either find the SM Higgs or tell us what the mechanism of EWSB truly is.

Looking back beyond the era of EWSB, all bets are off when it comes to the SM. Most strikingly, the origin of matter and energy in the cosmos cannot be explained within the SM at all. The most anthropically relevant component of this cosmic energy density -- the visible, baryonic matter component (about 5\%) -- could have been explained by the SM but it turns out the SM interactions fall short of what is required for such an explanation. The next biggest fraction corresponding to the cold dark matter (CDM) (about 25\%) has no particle candidate within the SM. Finally, the dark energy responsible for cosmic acceleration and comprising about 70\% of the cosmic energy density is the biggest mystery. While various BSM scenarios provide potentially feasible explanations for the baryonic and dark matter, our ideas about the dark energy remain the most speculative at present. For nuclear physicists, understanding the origin of the baryonic matter is quite important, as baryons\footnote{I include nuclei under this rubric} and their interactions with leptons and other hadrons are the bread and butter of the field.

The pre-EWSB era presents additional puzzles having to do with the unification of forces, stability of the electroweak scale, and neutrinos. If -- as many believe -- all forces were unified into a single interaction that included gravity at the end of the Big Bang, then the electroweak and strong couplings of the SM ought to meet at a common point when evolved to scales around 10$^{16}$ GeV. The problem is that they don't quite do so; there is something of a \lq\lq near miss" for grand unification. Consequently, one would need new physics to bring about unification, just as one does to explain the origin of matter. Generally speaking, the introduction of new physics tends to push the electroweak scale, given by the Higgs vev $v\approx 246$ GeV, up to higher scales. A larger value of $v$ would be problematic for the present SM universe because the Fermi constant that characterizes the strength of low-energy weak interactions is given by $G_F=(\sqrt{2}v^2)^{-1}$. Thus, larger the values of $v$ would lead to more feeble low-energy weak interactions, a sun that burns less brightly, and likely a different combination of light elements produced in Big Bang Nucleosynthesis than we observe in our universe. Clearly, new symmetries are required to preserve the relatively large value of $G_F$ in the presence of new physics needed to explain the origin of matter and bring about unification\footnote{One could get around the need for new symmetries if one allows for fine tuning.}.

Finally, neutrinos cause all kinds of consternation for the SM. The SM can be minimally extended to include Dirac neutrino mass terms with SU(2)$_L\times$U(1)$_Y$ right-handed (RH) neutrino fields, but the corresponding Yukawa couplings would have to be considerably smaller than for the other SM particles. This may, in fact, be what nature has given us, but many people believe a more natural explanation of the tiny neutrino masses arises if neutrinos are Majorana particles. In this case, the scale of neutrino mass is naturally given by $\sim m_e^2/M_{\rm new}$, where $M_{\rm new}$ is about 10$^{12}$ GeV. At present, however, we have no evidence that this scenario is correct\footnote{In many string-inspired models studied to date, it appears to be quite difficult to obtain Majorana mass terms or the minimal see-saw mechanism\cite{Giedt:2005vx,Langacker:2005pf}.}. Even more puzzling is the pattern of neutrino mixing. In contrast to the situation with quarks, where the mixing of weak eigenstates into mass eigenstates involves relatively small angles, just the opposite is true for neutrinos. Clearly, new physics is required to explain the differences in the patterns of mass generation or chiral symmetry-breaking for charged fermions and neutrinos. 

In the search for new symmetries that address these issues, the next high energy frontier is obviously the Large Hadron Collider. A second frontier -- the precision frontier -- now lies in the domain of low-energy studies that are being carried out by nuclear and atomic physicists. Information from the latter complements the former, as I hope to describe below. In doing so, I will divide the discussion in three parts: the origin of matter and EDM searches; precision studies of SM -allowed processes; and neutrinos. In doing so, I will not provide a comprehensive list of references, as space considerations preclude this possibility. More extensive reviews with reference to the literature can be found in Refs.~\cite{Erler:2004cx,Ramsey-Musolf:2006vr,Pospelov:2005pr,Severijns:2006dr,Herczeg:2001vk}

\section{EDMs and the Origin of Matter}

EDM searches are a particularly apt topic for this chiral dynamics meeting since the EDM operator is chiral odd. It is useful to write down the relevant SU(2)$_L\times$U(1)$_Y$ dimension six operators that involve the electric dipole couplings to the $W$ and $B$ gauge fields and include the relevant Higgs field insertions as needed for gauge invariance:
\begin{eqnarray}\label{eq:L6}
{\cal L}^{(6)}_{\rm CPV} & = & \frac{i\, g_1 d_u^B}{\Lambda^2} {\bar Q} \sigma_{\mu\nu}\gamma_5 B^{\mu\nu} {\tilde H} U + \frac{i\, g_1 d_d^B}{\Lambda^2} {\bar Q} \sigma_{\mu\nu} \gamma_5 B^{\mu\nu}  H D \\
\nonumber
& +&  \frac{i\, g_2 d_u^W}{\Lambda^2} {\bar Q} \sigma_{\mu\nu} \gamma_5 \tau^A W^{\mu\nu\, A} {\tilde H} U + \frac{i\, g_2 d_d^W}{\Lambda^2} {\bar Q} \sigma_{\mu\nu}\gamma_5 \tau^A W^{\mu\nu\, A}  H D+\cdots
\end{eqnarray}
where $Q$ is the LH quark doublet; $U$ and $D$ are RH quark singlets; ${\tilde H}=\epsilon_{ij} H^\ast_j$:\;  $\Lambda$ is a mass scale associated with the CP-violating (CPV) physics;  and the $+\cdots$ indicate other CPV interactions, such as those involving gluons. The expression for charged leptons is similar.  In Eq.~(\ref{eq:L6}). After EWSB, one obtains from these interactions the corresponding EDMs
\begin{eqnarray}
{\cal L}_{EDM} = -\frac{i\, d_u^\gamma}{2\Lambda} {\bar U}_L \sigma_{\mu\nu} F^{\mu\nu} U_R- \frac{i\, d_d^\gamma}{2\Lambda} {\bar D}_L \sigma_{\mu\nu} F^{\mu\nu} D_R-\frac{i\, d_\ell^\gamma}{2\Lambda} {\bar \ell}_L \sigma_{\mu\nu} F^{\mu\nu} \ell_R
\end{eqnarray}
where, for example, the EDM of the u-quark is given by
\begin{equation}
d_u^\gamma  = -\frac{\sqrt{2}\, v_u\left(c_W\, d_U^B+s_W\, d_U^W\right)}{\Lambda}\ \ \ .
\end{equation}
Note that the EDM operators are expressly chiral odd, as they involve one LH and one RH quark field. 

The present experimental bounds on EDMs of the electron and systems built out of quarks are quite stringent. In the case of the electron and neutron, for example, one has (for references to the experimental literature, see, {\em e.g.}, Refs.~\cite{Erler:2004cx,Ramsey-Musolf:2006vr,Pospelov:2005pr})
\begin{eqnarray}
\label{eq:EDMlim}
\left\vert d_e^\gamma/{\Lambda}\right\vert &<& 1.6\times 10^{-27}\ \ e-{\rm cm}\\
\left\vert {d_n^\gamma}/{\Lambda}\right\vert &<& 3.0\times 10^{-26}\ \ e-{\rm cm}
\nonumber
\end{eqnarray}
at 90\% confidence. The expectations for these EDMs within the electroweak sector of the SM lie many orders of magnitude below these values, but different BSM scenarios can lead to significantly larger EDMs. In the minimal supersymmetric standard model (MSSM), for example, one has for the electron at one-loop order
\begin{equation}
\label{eq:oneloopest}
\frac{d_e^\gamma}{\Lambda} \approx 5\times 10^{-25}\, \left(\frac{100\, {\rm GeV}}{{\tilde m}}\right)^2 \left[ \tan\beta\sin\phi_\mu - 0.05\sin\phi_A\right]\ e-{\rm cm}\ \ \ ,
\end{equation}
where $\phi_\mu$ and $\phi_A$ are CPV phases, ${\tilde m}$ is the mass of the supersymmetric particles (assuming their masses are degenerate), and $\tan\beta$ is the ratio of the vevs of the two Higgs doublets that arise in the MSSM. For $\tan\beta$ of ${\cal O}(1)$ and ${\tilde m}=100$ GeV, the bounds in Eq.~(\ref{eq:EDMlim}) imply that $\phi_\mu\lsim 3\times 10^{-3}$. 

From the standpoint of the cosmology, it turns out that this value of the CPV phase is too small to explain the matter-antimatter asymmetry within SUSY. The asymmetry itself is quite small and can be characterized by the baryon to photon entropy density
\begin{eqnarray}
\label{eq:ewb1}
Y_B\equiv \frac{n_B}{s} = 
\biggl\{
\begin{array}{cc}
(7.3\pm 2.5)\times 10^{-11}, & {\rm BBN }\\
(9.2\pm 1.1)\times 10^{-11}, & {\rm WMAP}
\end{array}
\end{eqnarray}
where \lq\lq BBN" and \lq\lq WMAP" indicate values derived from Big Bang Nucleosynthesis\cite{Eidelman:2004wy} and the cosmic microwave background \cite{Spergel:2003cb}, respectively. 

One can relax the EDM constraints while producing the value of $Y_B$ given in Eq.~(\ref{eq:ewb1})  by allowing the scalar superpartners of leptons and quarks become heavier -- of order a few TeV -- while keeping the masses of the gauge boson superpartners relatively light. In this scenario, the EDMs of elementary fermions are dominated by two-loop graphs, and CPV phases of ${\cal O}(1)$ are consistent with the experimental EDM bounds.  In order to obtain the observed value of $Y_B$ one would need superpartner masses and CPV phases consistent with an electron EDM of roughly $10^{-28}$ $e$-cm or larger (and similarly for the neutron). What makes the next several years particularly interesting from this standpoint is that new experiments are poised to search for  EDMs with precisely this magnitude or even small. Thus, one has some chance of either seeing the CPV that could explain the matter-antimatter asymmetry or ruling out conventional SUSY models as a mechanism for producing it. 

Making a robust connection between the results of future EDM experiments and the value of $Y_B$ requires careful analysis of the dynamics of baryon number generation during the electroweak phase transition (EWPT). Baryon number is produced by anomalous, topological transitions known as sphaleron processes. The effect of sphalerons live on the presence non-zero chiral charge density, and the latter is produced by CPV interactions of matter fields with the spacetime varying Higgs vevs at the boundary between regions of broken and unbroken electroweak symmetry. Thus, the CPV must be sufficiently effective to produce enough chiral charge to make the baryon number we observe today. Moreover, the phase transition must be strongly first order to ensure that as the region of broken electroweak symmetry expands, the sphalerons get sufficiently quenched that they cannot wash out the produced baryon number. In the SM, the lower bounds on the mass of the Higgs imply that a SM phase transition cannot be first order, but various BSM scenarios with extended Higgs sectors can produce such a strong first order phase transition. 

Although considerable theoretical progress has been made in performing refined computations of these effects (for the recent literature, see, {\em e.g.}, Ref.~\cite{Ramsey-Musolf:2006vr}), there remains considerable room for future theoretical progress.  The prospect of significantly more sensitive EDM searches provides powerful motivation for making such progress. A similar comment applies to computing EDMs within various BSM CPV scenarios. Clearly, considerations of chiral symmetry -- both through the EDMs themselves and through the dynamics of the electroweak phase transition -- are central to this problem of the origin of matter.

\section{Precision Electroweak Probes of New Symmetries}

Chiral symmetries can be a similarly interesting consideration when considering precision measurements of observables that are not suppressed in the SM. The presence or absence of tiny deviations from SM expectations can provide important clues for BSM physics. Perhaps the most widely-known recent example is the anomalous magnetic moment of  the muon, where many argue one now sees a deviation of nearly three standard deviations from the SM prediction. As with the EDM, the magnetic moment is a chiral-odd operator. The pure QED contributions to the corresponding operator coefficient, $a_\mu=(g_\mu-2)/2$, have been computed to high precision, as have one-loop electroweak contributions. The effects of strong interactions that enter the hadronic two-loop vacuum polarization and three-loop hadronic light-by-light contributions have proven more challenging theoretically. Assuming the relevant theoretical uncertainties are under control, the deviation from the SM expectation could be a signature of supersymmetric loop effects if $\tan\beta$ is large. 

Chiral-odd BSM effects can also enter low-energy weak interaction observables at an observable level. In the case of weak decays of hadrons, such as neutron $\beta$-decay and pion-decay, the low-energy semileptonic interaction can be described by a dimension six four fermion Lagrangian\cite{Profumo:2006yu}
\begin{equation}
\label{eq:leffbeta}
{\cal L}^{\beta-\rm decay} = - \frac{4 G_\mu}{\sqrt{2}}\ \sum_{\gamma,\, \epsilon,\, \delta} \ a^\gamma_{\epsilon\delta}\, 
\ {\bar e}_\epsilon \Gamma^\gamma \nu_e\, {\bar u} \Gamma_\gamma d_\delta
\end{equation}
where the $a^\gamma_{\epsilon\delta}$ coefficients are determined by the SM and its possible extensions. At tree-level in the SM, $a^V_{LL}=V_{ud}$ with all others being zero. There exist several equivalent representations of the low-energy effective semileptonic interaction\cite{Severijns:2006dr,Herczeg:2001vk}, but I prefer the form in Eq.~(\ref{eq:leffbeta}) because of its similarity to the muon decay effective Lagrangian. Note that none of these forms is invariant under the SM gauge symmetries and must, therefore, be used only for interactions taking place at energies well below the weak scale. 

As discussed elsewhere\cite{Severijns:2006dr,Herczeg:2001vk,Profumo:2006yu}, the study of $\beta$-decay correlations can probe for the existence of non $(V-A)\times(V-A)$ interactions appearing in Eq.~(\ref{eq:leffbeta}). Bounds on these interactions obtained from existing measurements can be found, {\em e.g.}, in Ref.~\cite{Severijns:2006dr}. Future, more precise probes may be obtained with studies of cold and ultracold neutrons at LANSCE, NIST, ILL, the SNS, and various other laboratories. From the standpoint of chiral symmetry, the scalar and tensor interactions in Eq.~(\ref{eq:leffbeta}) are interesting since they are chiral odd. Such operators can be generated in various BSM scenarios. In SUSY, for example, mixing between the superpartners of LH and RH fermions in loop graphs can give rise to the operators proportional to $a^S_{RR}$, $a^S_{RL}$, and $a^T_{RL}$. These operators generate $\beta$ energy-dependent contributions to the parity-violating correlation of the neutrino with the spin of the decaying nucleus (the \lq\lq $B$-term") as well as the so-called Fierz interference term (the \lq\lq $b$-coefficient") that also depends on the $\beta$ energy. Current limits on $b$ are at the $10^{-3}$ level while future measurements of the energy-dependence of $B$ at the $10^{-4}$ level may be achievable with cold or ultracold neutrons. 

Effects of this size could be generated in SUSY if the LH-RH first generation superpartner mixing is nearly maximal -- a situation that would be interesting since it would imply that the extra Higgs bosons in SUSY are too heavy to be seen at the LHC or that one must admit considerable fine-tuning to obtain appropriate electroweak symmetry-breaking. It is interesting to note that the mixing between LH and RH superpartners implies the explicit breaking of chiral symmetry in the scalar superpartner sector. The question is whether this breaking is small ({\em i.e.}, proportional to the fermion Yukawa couplings) for all the but the third generation sfermions -- as for the SM particles - or anomalously large. To date, I am not aware of explicit experimental tests of this possibility for the first generation quark and lepton superpartners. 

Chiral symmetry considerations are clearly important for discussions of pion decay as well as for $\beta$-decay. In the present context, it is interesting to consider the purely leptonic decays of the pion, from which we obtain the value of the pion decay constant, $F_\pi$, that is so important for the chiral dynamics of QCD. If we include the effects of electroweak radiative corrections in the SM as well as possible effects of new physics, the decay rate is given by (for reference to the literature, see, {\em e.g.}, Ref.~\cite{Ramsey-Musolf:2006vr})
\begin{eqnarray}
\label{eq:pionc}
\Gamma[\pi^+ \to \ell^+ \bar\nu_\ell (\gamma)] = {G_\mu^2 |V_{ud}|^2 \over 
  4\pi} F_\pi^2 m_\pi m_\ell^2 \left[ 1 - {m_\ell^2\over m_\pi^2} \right]&&
\\
\nonumber
\times  \left\{1+\left(2\left[{\Delta\hat r^A_\pi}-{\Delta\hat r}_\mu\right] +{\rm brem}\, \right)_{\rm SM} +2\left({\Delta\hat r^A_\pi}-{\Delta\hat r}_\mu\right)_{\rm new}\right\}&&
\end{eqnarray}
where ${\Delta\hat r^A_\pi}$ and ${\Delta\hat r}_\mu$ are corrections that come from SM  radiative corrections to the fundamental semileptonic and $\mu$-decay amplitudes, respectively (the \lq\lq SM " subscript) or from BSM physics (\lq\lq new" subscript) and \lq\lq brem" indicates the contributions from real photon radiation (required to keep the rate infrared finite). Taking just the SM contributions and the corresponding low-energy QCD related uncertainties in the SM radiative corrections as well a the experimental error in the rate, one obtains $F_\pi =92.4\pm 0.025 ({\rm expt}) \pm 0.25 ({\rm theory})$. Allowing for possible contributions from new physics can lead to further increases in the error. For example, if one allows for new tree-level SUSY interactions that violate lepton number conservation, present experimental constraints on these interactions leaves room for an additional $\sim 0.25\%$ uncertainty in the value of $F_\pi$ -- comparable to the present SM QCD uncertainty.

A well-known way to circumvent the largest QCD uncertainties and to actually use $\pi_{\ell 2}$ decays to probe BSM physics is to consider the ratio $R_{e/\mu}$ of the rates for decays to an electron-neutrino and muon-neutrino final state. The SM prediction for this quantity which tests \lq\lq lepton universality" is $R_{e/\mu}=(1.2352\pm 0.0005)\times 10^{-4}$ where the error is dominated by uncertainties in various low energy constants. This departure from unity follows simply from the different masses of the final state charged leptons. 
Experimentally, one finds 
\begin{eqnarray}
\label{eq:remuresult}
   {R_{e/\mu}^{\rm exp}\over R_{e/\mu}^{\rm SM}}=0.9966\pm 0.0030\pm 0.0004,
\end{eqnarray}
where the first error is experimental and the second theoretical (experimental references may be found in Refs.~\cite{Erler:2004cx,Ramsey-Musolf:2006vr}). A new generation of experiments are poised to measure $R_{e/\mu}$ with five to ten times smaller error bars, making the experimental uncertainty comparable to the theory error. Tests at this level could be quite interesting for SUSY, where lepton universality can be broken by differences in the lepton superpartner masses, leading to corrections as large as a few times $10^{-3}$. One could imagine further improvements in this experimental probe of \lq\lq slepton universality" by reducing the theoretical in the SM prediction -- clearly a task for chiral dynamics in QCD.

\section{Neutrinos and Chiral Symmetry}

The fact that neutrinos have small, but non-vanishing masses can have significant implications for other properties of neutrinos that would be forbidden in the presence of exact chiral symmetry. For example, the magnetic moments of neutrinos, which correspond to chiral odd operators, are constrained by the scale of neutrino mass and \lq\lq naturalness" considerations\cite{Bell:2005kz,Davidson:2005cs,Bell:2006wi}. The basic idea is quite simple. Insertions of the magnetic moment operator in loop graphs generates contributions to the neutrino mass operator, as both are chiral odd\footnote{Here, I focus on the case of Dirac neutrinos for simplicity.}. In order to avoid requiring large, \lq\lq un-natural" cancellations between these loop effects and tree-level contributions as needed to obtain the small scale of neutrino mass, the magnetic momenta operator coefficients cannot be too large. 

Model-independent constraints implementing this idea can be obtained by considering an effective Lagrangian containing SM fields and RH neutrino fields
\begin{equation}
\label{eq:leffnu}
{\cal L}_{\rm eff} = \sum_{n,j} \frac{C_j^n(\mu)}{\Lambda^{n-4}}\, {\cal O}_j^{(n)}(\mu) \  + {\rm h.c.}
\end{equation}
where $\Lambda$ is the mass scale associated with BSM physics and $j$ labels all operators of dimension $n\geq 4$. The $n=4$ operators as just those of the SM plus a neutrino Dirac mass term while Majorana mass terms appear at $n=5$. Magnetic moment operators for Dirac neutrinos appear at $n=6$ and have the gauge-invariant form 
\begin{eqnarray}
{\cal O}_{B}^{(6)} & = & g_{1}(\bar{L}\sigma^{\mu\nu}\widetilde{H})\nu_{R} B_{\mu\nu} \\
{\cal O}_{W}^{(6)} & = & g_{2}(\bar{L} \sigma^{\mu\nu}\tau^{a}\widetilde{H})\nu_{R} W_{\mu\nu}^{a}\ \ \ .
\end{eqnarray}
After EWSB, these operators generate a neutrino magnetic moment $\mu_\nu/\mu_B=-4\sqrt{2} (m_e v/\Lambda^2)[C_B^6(v)+C_W^6(v)]$. Radiative contributions to the neutrino mass operators and naturalness considerations lead to bounds on the coefficients $C_{B,W}$. For $\Lambda >> v$ the most stringent expectations arise from considering contributions to the $n=4$ neutrino mass operators and matching of the effective theory described by Eq.~(\ref{eq:leffnu}) onto the (unspecified) full theory. However, for $\Lambda$ not too different from $v$ mixing among the magnetic moment operators and the $n=6$ mass leads to bounds of comparable magnitude. 

From these arguments one expects the Dirac neutrino magnetic moments to be bounded above by
\begin{equation}
\left\vert\mu_\nu\right\vert/\mu_B \lsim 10^{-14}\times \left (m_\nu/1\, {\rm eV}\right)\ \ \ .
\end{equation}
These bounds are two or more orders of magnitude more stringent than the present experimental bounds on $\mu_\nu$. For Majorana neutrinos, the situation is more subtle. For $\Lambda\sim v$, the bounds on the transition moments are weaker than present experimental limits, while for $\Lambda\gsim 100$ GeV, the expectations are that the transition moments would be smaller than present direct constraints. Given the expected sensitivity of future neutrino magnetic moment searches, the discovery of a non-zero moment would imply that the neutrino is a Majorana particle and that the mass scale $\Lambda$ of the corresponding BSM physics is well below the standard see-saw scale of $\sim 10^{12}$ GeV. These conclusions could only be altered in specific models wherein these chiral symmetry-based expectations allow for considerably larger magnetic moments through the introduction of other mechanisms that protect $m_\nu$ from larger corrections or through the presence of non-SM mechanisms for generating charged lepton masses. Applications of these neutrino mass naturalness arguments to other processes can be found in Refs.~\cite{Prezeau:2004md,Erwin:2006uc}.

\section{Conclusions}

I hope to have convinced the reader that low-energy searches for BSM physics are an appropriate topic for a meeting on chiral dynamics. Chiral symmetries can play a significant role in both the computation of SM observables as well as in BSM scenarios that generate corrections to SM expectations. Indeed, the presence of (broken) chiral symmetries are a key element in explaining  the origin of matter and CPV in the early universe; the properties and interactions of neutrinos; and the weak decays of leptons and systems built from light quarks. The next several years promises to be an exciting time in the study of these phenomena, and we may expect reports of interesting experimental and theoretical developments at future chiral dynamics conferences.

\section{Acknowledgments} 
I would like to thank N. Bell, V. Cirigliano, J. Erler, R. Erwin, B. Filippone, M. Gorshteyn, B. Holstein, J. Kile, P. Langacker, C. Lee, W. Marciano, R. McKeown, B. Nelson, S. Profumo, S. Su, S. Tulin, P. Vogel, P. Wang, and M. Wise for useful conversations. This work was supported under U.S. Department of Energy contract DE-FG02-05ER41361 and National Science Foundation Award PHY-05556741.


\begin{thebibliography}{99}

\bibitem{lepewwg07} http://lepewwg.web.cern.ch/LEPEWWG

\bibitem{Ramsey-Musolf:2006ur}
  M.~J.~Ramsey-Musolf,
  AIP Conf.\ Proc.\  {\bf 842} (2006) 661
  [arXiv:hep-ph/0603023].

\bibitem{Ramsey-Musolf:2006ik}
  M.~J.~Ramsey-Musolf,
  arXiv:nucl-th/0608035.

\bibitem{Giedt:2005vx}
  J.~Giedt, G.~L.~Kane, P.~Langacker and B.~D.~Nelson,
  Phys.\ Rev.\ D {\bf 71}, 115013 (2005)
  [arXiv:hep-th/0502032].
  
\bibitem{Langacker:2005pf}
  P.~Langacker and B.~D.~Nelson,
   ``String-inspired triplet see-saw from diagonal embedding of SU(2)L in
  Phys.\ Rev.\ D {\bf 72}, 053013 (2005)
  [arXiv:hep-ph/0507063].

\bibitem{Erler:2004cx}
  J.~Erler and M.~J.~Ramsey-Musolf,
  Prog.\ Part.\ Nucl.\ Phys.\  {\bf 54} (2005) 351
  [arXiv:hep-ph/0404291].

\bibitem{Ramsey-Musolf:2006vr}
  M.~J.~Ramsey-Musolf and S.~Su,
  arXiv:hep-ph/0612057.

\bibitem{Pospelov:2005pr}
  M.~Pospelov and A.~Ritz,
  Annals Phys.\  {\bf 318}, 119 (2005)
  [arXiv:hep-ph/0504231].
  

\bibitem{Severijns:2006dr}
  N.~Severijns, M.~Beck and O.~Naviliat-Cuncic,
  arXiv:nucl-ex/0605029.
  

\bibitem{Herczeg:2001vk}
  P.~Herczeg,
  Prog.\ Part.\ Nucl.\ Phys.\  {\bf 46} (2001) 413.


\bibitem{Eidelman:2004wy}
  S.~Eidelman {\it et al.}  [Particle Data Group],
  Phys.\ Lett.\ B {\bf 592} (2004) 1.

\bibitem{Spergel:2003cb}
  D.~N.~Spergel {\it et al.}  [WMAP Collaboration],
  Astrophys.\ J.\ Suppl.\  {\bf 148} (2003) 175
  [arXiv:astro-ph/0302209].

\bibitem{Profumo:2006yu}
  S.~Profumo, M.~J.~Ramsey-Musolf and S.~Tulin,
  arXiv:hep-ph/0608064.

\bibitem{Bell:2005kz}
  N.~F.~Bell, V.~Cirigliano, M.~J.~Ramsey-Musolf, P.~Vogel and M.~B.~Wise,
  Phys.\ Rev.\ Lett.\  {\bf 95} (2005) 151802
  [arXiv:hep-ph/0504134].

\bibitem{Davidson:2005cs}
  S.~Davidson, M.~Gorbahn and A.~Santamaria,
  Phys.\ Lett.\ B {\bf 626} (2005) 151
  [arXiv:hep-ph/0506085].

\bibitem{Bell:2006wi}
  N.~F.~Bell, M.~Gorchtein, M.~J.~Ramsey-Musolf, P.~Vogel and P.~Wang,
  Phys.\ Lett.\ B {\bf 642} (2006) 377
  [arXiv:hep-ph/0606248].

\bibitem{Prezeau:2004md}
  G.~Prezeau and A.~Kurylov,
  Phys.\ Rev.\ Lett.\  {\bf 95} (2005) 101802
  [arXiv:hep-ph/0409193].


\bibitem{Erwin:2006uc}
  R.~J.~Erwin, J.~Kile, M.~J.~Ramsey-Musolf and P.~Wang,
  arXiv:hep-ph/0602240.

\end{thebibliography}
\end{document}